\documentclass[twocolumn, superscriptaddress,  floatfix, unsortedaddress, aps, prl, showpacs, amssymb,amsmath]{revtex4}
\usepackage{graphics}
\usepackage[T1]{fontenc}
\usepackage[latin1]{inputenc}

\begin{document}

\title{Strain-dependent localization, microscopic deformations, and
macroscopic normal tensions in model polymer networks}

\author{Carsten Svaneborg}
\email{zqex@mpipks-dresden.mpg.de}
\affiliation{Max-Planck-Institut f\"ur Polymerforschung, Postfach 3148, D-55021
Mainz, Germany}
\affiliation{Max-Planck-Institut f\"ur Physik komplexer Systeme,
N\"othnitzer Str. 38, Dresden, Germany}
\author{Gary S. Grest}
\affiliation{Sandia National Laboratories, Albuquerque, NM 87185, USA}
\author{Ralf Everaers}
\affiliation{Max-Planck-Institut f\"ur Physik komplexer Systeme,
N\"othnitzer Str. 38, Dresden, Germany}

\begin{abstract}  
We use molecular dynamics simulations to investigate the microscopic and
macroscopic response of model polymer networks to uniaxial elongations.  By
studying networks with strands lengths ranging from $N_s=20$ to $200$ we
cover the full crossover from cross-link to entanglement dominated
behavior. Our results support a recent version of the tube model which
accounts for the different strain dependence of chain localization due to
chemical cross-links and entanglements.
\end{abstract}


\maketitle

Cross-linking a melt of linear precursor chains leads to a polymer network
which macroscopically behaves as a (viscoelastic) solid and which is
microscopically characterized by a complex, quenched, random connectivity and
topology~\cite{deam76}. Over the past sixty years a large variety of theories
of rubber elasticity has been put
forward~\cite{treloar75,heinrich88,edwards88v}, and there is a corresponding
body of rheological literature devoted to comparing and testing the proposed
stress-strain relations, see e.g.
\cite{Gottlieb_pol_83,Urayama}. Only recently, neutron scattering experiments
\cite{StraubeUrban,StraubeRichter} and computer simulations
\cite{duering94,everaers95,everaers99,grest00,Sommer2002} have begun to provide
detailed microscopic information. 
Attempts to quantitatively correlate the microscopic and macroscopic
response to strain were so far restricted to 
idealized model polymer networks with diamond lattice
connectivity~\cite{everaers95,everaers99}.

In this Letter we report a comprehensive set of results from computer
simulations of randomly end-linked model polymer networks
\cite{duering94,grest00} under elongational strain.  
In our data analysis we follow the logic of most statistical
mechanical theories of rubber elasticity
\cite{heinrich88,edwards88v}, i.e. we  relate the localization of different
parts of a polymer network to strain-induced microscopic deformations and the
macroscopic elastic response \cite{treloar75,doi86}.  We focus on
theories~\cite{deam76,WarnerEdwards,heinrich88,rubinstein97,read97,everaers98,ME} based on Edwards' tube model~\cite{edwards67}, because the underlying
ideas are conceptually relatively simple, (almost) completely worked out, and
closely related to most modern theories of polymer
rheology \cite{doi86,mcleish02}. Nevertheless, a key ingredient of the model,
the strain and stand length dependence of the phenomenological tube diameter,
remains controversial~\cite{read_prl,straube_comment}.  Our data
support a recent generalization of the tube model~\cite{ME} which accounts for
the different character of chain localization by
cross-links~\cite{WarnerEdwards,read97} and
entanglements~\cite{heinrich88,rubinstein97,everaers98}.

We used extensive Molecular Dynamics (MD) simulations to study the behavior
of bead-spring polymer melts under uni-axial, volume-conserving elongation.
The polymer model has two interactions that represent bonds and excluded
volume, respectively. Interaction parameters are chosen to ensure
conservation of the topological state.  The details of the simulation
methodology can be found in the literature, see e.g.
\cite{duering94,grest00}. The present Letter is based on simulations of 
end-linked networks of $M\times{}N_s=$ $5000\times{}20$, $1000\times{}35$,
$2500\times{}100$, and $3000\times{}200$ where $M$ denotes the number of
strands and $N_s$ strand length. 
The state of the networks can be accurately
characterized: the gel faction is $>99\%$, $>91\%$ of the network is
elastically active, and the fraction of four-functional crosslinkers is
$>78\%$, except for $62\%$ in the case of the $3000\times200$. Each network
was simulated at several elongations
($\lambda_x=$$\lambda_{\parallel}=\lambda $,
$\lambda_y=$$\lambda_z=$$\lambda_{\perp }=$$1/\sqrt{\lambda } $). The networks
were successively strained. After each strain increment the network was
equilibrated, and configurations were samped for more than $30$ ($2$) Rouse
times of the strands for the two short (long) strand networks,
respectively. To reduce finite chain length effects the maximal elongations
were limited to $\lambda=2$ ($\lambda=4$) for short (long) strand networks,
respectively. Elastic properties were obtained by sampling the deviatoric part
of the microscopic virial tensor defined as 
$\sigma_{\alpha\beta}=\left\langle \sum_{ij} F_{ij,\alpha} r_{ij,\beta}
\right\rangle/V$, where the sum is over all pairs $i,j$ of interacting beads,
$\alpha$, $\beta$ are Cartesian indices, and $F$, $r$ and $V$ denote forces,
separations and the volume, respectively.  The normal tension is defined as
$\sigma_T=\sigma_{xx}-(\sigma_{yy}+\sigma_{zz})/2$. In the following we
present our simulation results together with a brief outline of the
theoretical background.
\begin{figure}[t]
{\centering \resizebox*{1\columnwidth}{!}{\rotatebox{0}{\includegraphics{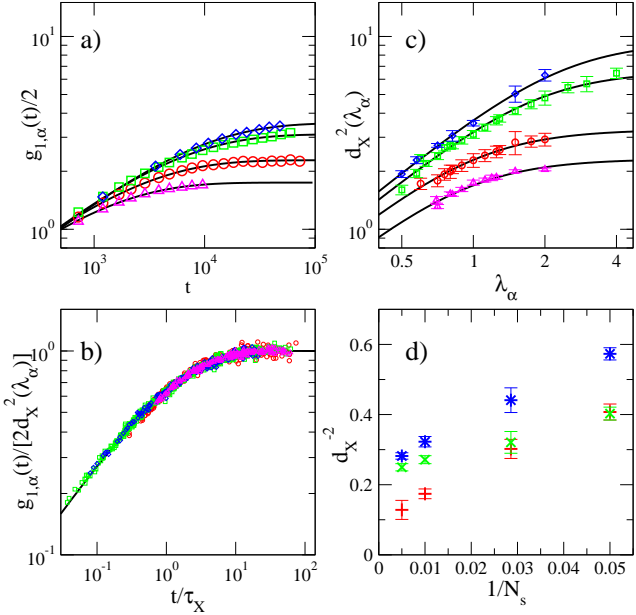}}} \par{} }
\caption{\label{fig:xlocalization}
(color online)
Strain-dependent localization of cross-links for $N_s=20$ (magenta
$\triangle$), $N_s=35$ (red $\circ$), $N_s=100$ (green $\Box$), and $N_s=200$ 
(blue $\diamond$).
(a) Cross-link mean-square displacements as a function of time for the
unstrained networks. Lines are fits of
$2 d^2_X(\lambda_\alpha)[1-\exp(-\sqrt{t/\tau_X})]$ which we use to
extract the tube diameters.
(b) Rescaled cross-link mean-square displacements for all
$\{N_s,\lambda_\alpha\}$ in comparison to the fit function.
(c) Strain dependence of the tube diameter.
The lines are fits of Eq.~(\protect\ref{eq:d}).
(d) Strand length dependence of crosslinker confinement
$d_X$ (blue $*$), $d_{X,A}$ component (red $+$), and $d_{X,B}$
component (green $\times$).
}
\end{figure}

With the tube model~\cite{edwards67} Edwards proposed a tractable
simplification of the complicated many-body problem of a randomly cross-linked
and entangled polymer network.  The key idea is that vulcanization leads to a
permanent {\em localization} of the precursor chains (or, equivalently, long,
randomly chosen paths through the network) in tube-like regions along the
coarse-grained chain contours. 
In the following $d(\lambda_\alpha)$ denotes the $\alpha$'th Cartesian
component of the tube diameter, which depends only on the corresponding strain
component $\lambda_\alpha$, hence each elongation provides two tube diameters
$d(\lambda_\parallel)$ and $d(\lambda_\perp)$. The zero-strain tube diameter
$d(\lambda_\alpha=1)$ is abbreviated $d$.
The simplest measure of the tube diameter is the width,
$d_X(\lambda_\alpha)$, of the thermal fluctuations of chemical cross-links
around their average positions (Fig.~\ref{fig:xlocalization}). Our simulation
runs are long enough to allow for a reliable extraction of $d_X$ from the
components of the mean-square displacements
$g_{1,\alpha}(t)=\langle\left[r_{\alpha}(t)-r_{\alpha}(0)\right]^2\rangle$,
where the average is restricted to four-functional cross-links (Figs.~\ref{fig:xlocalization}a and
b). 

The key issue is the non-trivial dependence of the extracted tube diameters on
strain and strand length (Fig.~\ref{fig:xlocalization}c and d).  From a
theoretical point of view, the situation is relatively clear in the
hypothetical case of ``phantom networks'' which consist of non-interacting
Gaussian polymer chains \cite{james43}. The corresponding tube theory by
Warner and Edwards (WE)~\cite{WarnerEdwards} uses an isotropic,
strain independent tube diameter $d_{X,A}(\lambda_\alpha) = d_{X,A}$, where
$d_{X,A}\sim b\sqrt{N_s}$ is of the order of the root-mean-square extension of
the network strands. The description of polymer networks as phantom networks
becomes inappropriate, if the length of the network strands approaches the
melt entanglement length, $N_e$ \cite{doi86,mcleish02}. In particular, the tube diameter should become {\em
independent} of strand length in the limit of very long strands where
$d_{X,B}\sim b\sqrt{N_e}$. The strain dependence of entanglement dominated
confinement is a subtle point. Empirical
evidence~\cite{StraubeRichter,everaers99} and theoretical
arguments~\cite{ronca75,heinrich88,rubinstein97,everaers98} support the tube
theories by Heinrich and Straube (HS)~\cite{heinrich88} and by Rubinstein and
Panyukov (RP)~\cite{rubinstein97} which predict an anisotropic, strain
dependent tube diameter of $d_{X,B}(\lambda_\alpha)=\sqrt{\lambda_\alpha}d_{X,B}$.

\begin{figure}[t]
{\centering \resizebox*{\columnwidth}{!}{\rotatebox{0}{\includegraphics{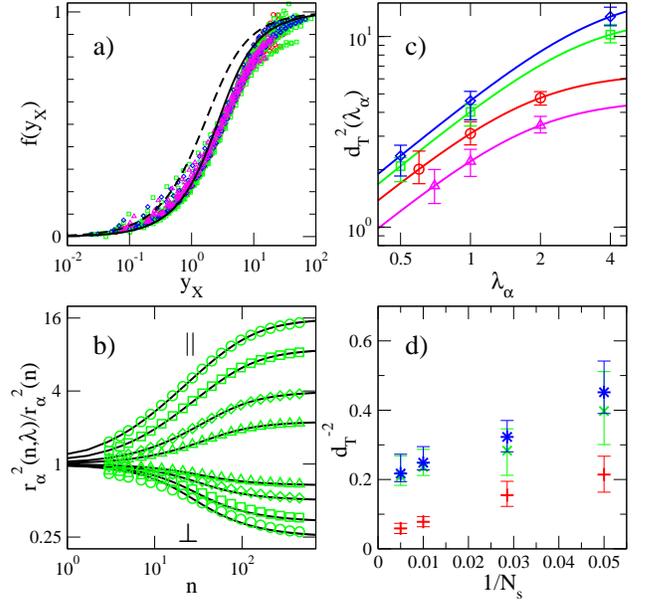}}} \par{} }

\caption{\label{fig:microdef}(color online)
  Length scale dependent microscopic deformation response (symbols as in 
  Fig.~\protect\ref{fig:xlocalization}).
  (a) Scaling plot of the degree of affineness of the microscopic deformations
      using the tube diameters from Fig.~\protect\ref{fig:xlocalization}c. The lines are
      $f_A$ (dashed) and $f_B$ (solid).
  (b) Parallel and perpendicular microscopic deformations for the
  $2500\times100$ system for strain $\lambda=1.5,2,3,4$ (symbols).
      Lines are fit of Eq.~(\protect\ref{eq:msd}).
  (c) Strain dependence of the tube diameters. Symbols indicate the error
      bars and the range of $\lambda_\alpha$ covered by the data.
  (d) Strand length dependence of tube confinement $d_T$ (blue $*$),
      $d_{T,A}$ component (red $+$), and $d_{T,B}$ component (green $\times$).
}
\end{figure}
Instead of combining an entanglement tube model with a classical
theory of rubber elasticity in an ad-hoc fashion, the recently
introduced double tube theory by Mergell and Everaers (ME)~\cite{ME} accounts
for the simultaneous presence of both effects with the help two correlated and
additive confining potentials of the WE and the HS-RP type, respectively.
The ME theory predicts
\begin{equation}\label{eq:d}
\frac1{d^{4}_{T}(\lambda_\alpha)}=\frac1{d^{4}_{A}}+\frac1{\lambda_\alpha^2 d^{4}_{B}}\ ,
\end{equation}
with an effective zero-strain tube diameter $d_T^{-4}=d_A^{-4}+d_B^{-4}$.
As indicated by the solid lines in Fig.~\ref{fig:xlocalization}c, the individual
data sets are well described by the functional form of Eq.~(\ref{eq:d}). 
Moreover, the strand length dependence of the fitted cross-link and
entanglement tube diameters $d_{X,A}$ and $d_{X,B}$ is in good agreement
with the arguments underlying the double tube theory (Fig.~\ref{fig:xlocalization}d).
It is worth noting the introduction of crosslinkers cause extra topological
constraints, hence the observed narrowing of the entanglement tube for $N_s=20$.

In the second part of our analysis, we consider microscopic deformations. In
rubber-like materials, macroscopic deformations affect distances between
neighboring monomers only weakly, while distances between distant monomers
change affinely with the macroscopic strain.  Following the logic of the tube
model, we focus on the length scale dependent deformations of non-reversal
random walk paths through the network. As a probe for microscopic deformations
we utilize the mean-square distances $r^2_\alpha(n,\lambda_\alpha)$ between
pairs of beads as a function of their chemical distance $n$. In a Gaussian
theory $r^2_\alpha(n,\lambda_\alpha)$ fully specifies the microscopic 
conformations. In the absence of strain
$r^2_\alpha(n)= b^2 n$, where $b^2$ is one Cartesian component of the
mean-square segment length for a precursor chain. End-linking is known not
to change the segment length compared to the precursor melt \cite{duering94}.
The crossover to affine deformations can be characterized by a dimensionless
function~\cite{ME}
\begin{equation}\label{eq:A}
f = \frac{ r^2_\alpha(n,\lambda_\alpha) - r^2_\alpha(n)}
{ (\lambda _{\alpha }^{2}-1) r^2_\alpha(n) }, 
\end{equation}
with $0<f<1$. According to the tube model the crossover length should be of the order of
$d_X$, i.e. $f$ should be a function of $y_X=r^2_\alpha(n)/[2d^{2}_{X}]$.
Fig.~\ref{fig:microdef}a verifies the interdependence of localization and
microscopic deformations, which is an essential element of the tube model. Note
that the plot contains {\em all} available data sets
($\{N_s,\lambda_\alpha\}$).

Theories based on the tube model make explicit predictions for the functional
form of $f$ (see the lines in Fig.~\ref{fig:microdef}a):
$f_{A}(y)=1+(\exp(-y)-1)/y$ \cite{WarnerEdwards} and $f_{B}(y)=1+0.5\exp
(-y)+1.5(\exp (-y)-1)/y$ \cite{ME} were obtained in the limits of cross-link
and entanglement dominated confinement, respectively. For the general case, the
double tube model \cite{ME} predicts
\begin{equation}\label{eq:msd}
f(y) = 
f_{A}(y)+
  \frac{ d_{T,A}^4 }{d_{T,B}^4\lambda _{\alpha }^{2}+d_{T,A}^4 }
\left[ f_{B}(y)-f_{A}(y)\right], 
\end{equation}
where $ y=r^2_\alpha(n)/[2 d^{2}_{T}(\lambda_{\alpha })]$ with
$d_T(\lambda_\alpha)$ given by Eq.~(\ref{eq:d}). 

We determined $d_{T,A}$ and $d_{T,B}$ for each system by fitting
Eq.~(\ref{eq:msd}) to the sampled $r^2_\alpha(n,\lambda_\alpha)$ for all
values of $\lambda_\alpha$ simultaneously \footnote{All parallel and perpendicular components
were fitted simultaneously, and logarithmically distributed chemical distances
were used to avoid systematic errors due to a large number of strongly
correlated points at large distances. The statistical error was estimated by a
block analysis of the path ensemble. A systematic error was estimated by
fitting the data restricted to $y\in [0.1:1]$ and $y\in [1:10]$ separately.}.
A typical result is shown in Fig. \ref{fig:microdef}b.  The effective
strain-dependent tube diameters are shown in Fig.\ref{fig:microdef}c, the two
components $d_{T,A/B}(N_s)$ are plotted in Fig.\ref{fig:microdef}d.  As
expected from the scaling plot Fig.~\ref{fig:microdef}a, there is qualitative
agreement between the tube diameters $d_{X}$ extracted from the cross-link
fluctuations and $d_{T}$ inferred from the analysis of the microsopic
deformations.  In particular, we observe a similar crossover to entanglement
dominated confinement for long strands.  Not surprisingly, there is no
quantitative agreement between the two measures of the tube diameter.  Tube
models of rubber elasticity do not distinguish between middle monomers of
network strands (which are free to slide along the entanglement tube) and
cross-linkers (which are not)~\cite{duering94}. In fact, the theories
discussed in this Letter employ a simple harmonic localization potential which
suppresses reptation-like motion.

\begin{figure}[t!] {\centering\resizebox*{1\columnwidth}{!}{
\rotatebox{270}{\includegraphics{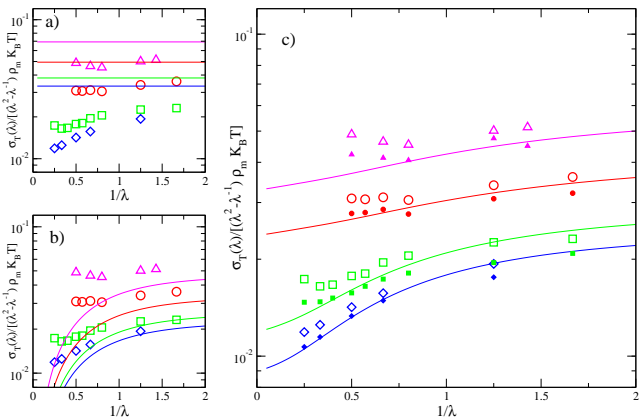}} } \par{}} 
\caption{\label{fig:MacroStress}(color online) Mooney-Rivlin plot of normal
tensions as a function of strain:  normal tensions sampled using the virial
tensor (large symbols as in Fig.~\ref{fig:xlocalization}), and ``Gaussian''
normal tensions $\sigma_{T,{\mathrm Gauss}}$ (small filled symbols) vs.
theoretical predictions based on the analysis of the microscopic deformations (
Fig.~\ref{fig:microdef}) (a) WE crosslinker tube theory, (b) HS-RP entanglement
tube theory, (c) ME theory. Symbols and lines have a 10\% and 30\% error,
respectively. } \end{figure}

In the third part of our analysis we consider the macroscopic elastic response
of our networks. Fig.~\ref{fig:MacroStress} shows the strain-dependence of the
sampled normal tensions in the form of a standard Mooney-Rivlin
plot~\cite{treloar75}. This representation is commonly used to emphasize
deviations from the classical stress-strain relation
$\sigma_T(\lambda)\propto(\lambda^2-\lambda^{-1})$.  Similarly to
experiments~\cite{treloar75}, we find
that these deviations are more pronounced for entanglement dominated systems
and that the shear modulus decreases with increasing strand length.

In polymer physics \cite{treloar75,doi86} stresses are usually derived from chain
conformations by dividing the chains into segments of length $n$. Segments are assumed
to behave as independent entropic springs with spring constant $k_BT/r^2_\alpha(n)$.
For a given monomer density $\rho_m$, the segment density is $\rho_s(n)=\rho_m/n$
\footnote{Treating the stress contributions of the elastically active and inactive
parts of the network separately changes $\sigma_{T,{\mathrm Gauss}}(\lambda)$ by less
than 5\%, as a consequence $\rho_m$ has not corrected to account for this effect.}. Normal
tensions derived from the virial tensor for this mesoscopic polymer model take a
simple form involving only those ratios of mean-square internal distances which we
have plotted in Fig.~\ref{fig:MacroStress}c:
\begin{equation}\label{eq:sigman}
\sigma_{T,{\mathrm Gauss}}(\lambda)= k_B T \lim_{n\rightarrow0} \rho_s(n) \left[ 
\frac{r^2_\parallel(n,\lambda_\parallel)}{r^2_\alpha(n)}
-\frac{r^2_\perp(n,\lambda_\perp)}{r^2_\alpha(n)} \right]\ ,
\end{equation}
There are
considerable subtleties in comparing the virial tensor calculated from the full
microscopic interactions to the ``Gaussian'' normal tensions~\cite{gao1}.
Fig.~\ref{fig:MacroStress}c shows that in the present case both quantities agree
within the statistical error even though $\sigma_{T,{\mathrm
Gauss}}(\lambda)/\sigma_{T}(\lambda)\approx 90\%$. In contradiction to the arguments
put forward in Ref.~\cite{Sommer2002}, we take the good agreement as quantitative
evidence, that the tube represention of the network conformation in terms of {\em
linear} paths through the network properly accounts for the relevant microscopic
deformations.

The final step of our analysis is the {\em parameter-free} comparison in
Fig.~\ref{fig:MacroStress}c of the measured stress-strain relations to the
prediction of the ME theory~\cite{ME}
\begin{eqnarray}
\label{eq:sigma} \sigma _{T}(\lambda) &=& 
(\lambda_\parallel^{2}-1)g(\lambda_\parallel )-
(\lambda_\perp^2-1)g(\lambda_\perp)\\ 
g(\lambda_\alpha)&=&
\frac{\rho_{m}k_{B}T}{8}
\frac{b^2}{d_{T}^2(\lambda_\alpha)}
\frac{d_{T,A}^4 + 2 \lambda_\alpha^2 d_{T,B}^4 }
{d_{T,A}^4 + \lambda_\alpha^2 d_{T,B}^4  } \ ,\nonumber
\end{eqnarray} 
where we use the values of $d_{T,A}$ and $d_{T,B}$ from our fits of
the microscopic
deformations. The measured and the inferred normal tensions agree within the
statistical error. To demonstrate the importance of the proper treatment of
the deformation dependence of the confinement, we have also calculated
stress-strain curves from the WE theory~\cite{WarnerEdwards}
($d_T(\lambda_\alpha) = d_T\Leftrightarrow d_{T,B}=\infty $) and the HS-RP
theory~\cite{heinrich88,rubinstein97} ($d_T(\lambda_\alpha) =
\sqrt{\lambda_\alpha} d_T\Leftrightarrow d_{T,A}=\infty $). The comparison in 
Figs.~\ref{fig:MacroStress}a and b shows that attempts along
these lines are restricted to estimates of the order of magnitude of the
elastic response.

To summarize, we have used computer simulations to determine strain-dependent
localization, length scale dependent microscopic deformations and macroscopic
stresses in model polymer networks. Fig.~\ref{fig:microdef}a directly
validates Edward's original tube concept: on a scaling level the same length
scale characterizes the spatial localization of chemical crosslinks and the
crossover from local liquid-like to global solid-like behavior of the
microscopic chain deformations. Closer inspection
(Figs.~\ref{fig:xlocalization}/\ref{fig:microdef} c,d) reveals the different
character of (and the crossover between) cross-link and entanglement dominated
confinement anticipated by the double tube model. In particular, Figs.~\ref{fig:microdef} and
\ref{fig:MacroStress} show that this model can {\em simultaneously}
describe microscopic and macroscopic aspects of the response of our model networks to
strain by accounting for the strain dependence of the (effective) tube
diameter.  While there clearly remain open questions concerning longitudinal
fluctuations in the entanglement
tube~\cite{edwards88v,Terentjev,Rubinstein2002}, we would like to emphasize
that our simulations provide a substantially improved empirical basis for 
addressing these and similar problems in the controlled development of statistical mechanical
theories of rubber elasticity. Currently, we are analysing the strain
dependence of the primitive path mesh~\cite{PPA} in an attempt to
systematically link the phenomenological tube model to the microscopic
connectivity and topology of our model networks.  Finally, we note that our
simulations can also help to validate critical steps in the data analysis of
(scattering) experiments addressing these issues.

\begin{acknowledgments} We greatfully acknowledge helpful discussions and a
long-standing collaboration with K. Kremer. We are particularly grateful to D.
Heine for contributing data for $3000\times200$ networks.  Sandia is a
multiprogram laboratory operated by Sandia Corporation, a Lockheed Martin
Company, for the United States department of Energy's National Nuclear
Security Administration under contract de-AC04-94AL85000.
\end{acknowledgments}

\end{document}